\documentclass[12pt]{article}

\usepackage[margin = 1.5in]{geometry}
\usepackage{amsmath}
\usepackage{amsfonts}
\usepackage{dsfont}
\usepackage{graphicx,psfrag,epsf}
\usepackage{color}
\usepackage{enumerate}
\usepackage{natbib}
\usepackage{hyperref}
\usepackage{caption}
\usepackage{subcaption}
\usepackage{float}
\usepackage{booktabs}
\usepackage{algorithm}
\usepackage{algpseudocode}

\allowdisplaybreaks

\newcommand{\blind}{1}

\newcommand{\V}{\mathcal{V}}

\newcommand{\rY}{Y^{(r)}}
\newcommand{\ry}{y^{(r)}}

\newcommand\numberthis{\addtocounter{equation}{1}\tag{\theequation}}

\begin{document}

\def\spacingset#1{\renewcommand{\baselinestretch}%
{#1}\small\normalsize} \spacingset{1}

\if1\blind
{
  \title{\bf Estimating Global Network Density from Snowball Samples}
  \author{Nurzhan Sapargali\\
    Department of Statistics, University of Munich\\
    and \\
    Sergio Buttazzo \\
    Department of Statistics, University of Munich \\
    and \\
    Göran Kauermann \\
    Department of Statistics, University of Munich}
  \maketitle
 } \fi

\if0\blind
{
  \bigskip
  \bigskip
  \bigskip
  \begin{center}
    {\LARGE\bf Estimating Global Network Density from Snowball Samples}
\end{center}
  \medskip
} \fi

\bigskip

\begin{abstract}
Many analyses aim to estimate a global property of a population network, such as its density, yet only a subsample of the network is observed. Ignoring the sampling scheme, and treating the samples as complete networks leads to biased inference.
Although this selection problem is generally intractable, we show that it admits an exact solution for snowball samples drawn from Erd\H{o}s--R\'enyi networks.
We derive the exact likelihood of such a sample, showing that it forms a curved exponential family in the edge probability with a low-dimensional sufficient statistic.
We obtain the maximum likelihood estimator that accounts for the sampling design, alongside valid confidence intervals constructed by inverting a test based on the exact sampling distribution, via Monte Carlo simulation.
Simulations show that this estimator remains unbiased even when observing as little as 0.1\% of the vertices, with nominal confidence interval coverage.
We further assess robustness to departures from the Erd\H{o}s--R\'enyi assumption using scale-free and small-world synthetic networks, as well as a real Facebook ego network.
We find that, while the corrected estimator no longer attains unbiasedness under these departures, it still substantially outperforms the naive alternative throughout.
\end{abstract}


\spacingset{1.45}

\section{Introduction}
\label{sec:intro}
Sample data are often obtained through mechanisms that are not indifferent to the values under study: whether a unit is observed, and which of its quantities are recorded, can depend on the very quantities the analysis seeks to estimate.
Treating such data as though they arose from an indifferent design can lead to systematically biased inference.
A substantial survey-sampling literature addresses when, and to what extent, the design by which a sample was selected can be disregarded once the data are in hand.
\citet{basu1969role} showed that a maximal sufficiency reduction always exists for a survey sampling model, arguing that inference should not depend on the design used to draw the sample.
\citet{scott1973survey} showed that this irrelevance holds in a Bayesian analysis only when sampled units retain identifying labels and the design is non-informative; without labels, the design contributes information of its own, except when an exchangeable prior is combined with interest restricted to symmetric functions of the data.
\citet{rubin1976inference} placed these ideas within a general theory of missing data, showing that sampling is a particular instance of a selection mechanism, and gave exact conditions under which such a mechanism can be ignored for sampling-distribution, likelihood, or Bayesian inference.
\citet{sugden1984ignorable} argue that these conditions are more fragile than they first appear: a design that is ignorable when fully known may become informative once only partial information about it is available to the analyst.
Together, this body of work establishes that likelihood-based inference can, under identifiable conditions, proceed correctly by conditioning on exactly what was observed and how it came to be observed.

Networks pose a distinctive instance of this problem.
When the population of interest is a network, the units available for observation are frequently discovered through the very relations under study, particularly when the population is hidden or difficult to enumerate directly \citep{thompson2000model}.
Link-tracing designs, in which respondents identify their contacts and those contacts are in turn added to the sample, are a practical and often unavoidable way of obtaining such data. In such settings, the observed network is only a subsample of a much larger population network, and natural targets of inference are global properties of that population, in particular its density. The density is a primary descriptive of network connectivity and affects, for instance, the average degree in the population. Estimating it from a link-traced subsample is therefore a practically relevant problem in its own right. It is also a hard one: the density of the sampled subnetwork is a poor proxy for the full population density, because the vertices reached by the sampling scheme are systematically more connected than a typical vertex, and because the size of the sample itself carries information about the density.

\citet[Definition 3.2]{crane2018network} formalizes the resulting challenge through \emph{consistency under subsampling}: a sampling scheme is consistent if the distribution induced on the observed network coincides with the distribution assigned to a network of that size under the population model.
This holds automatically for ordinary selection, where a fixed number of vertices is drawn independently of the graph's edge structure, but typically fails for link-tracing designs, since which vertices are included, and how many, depends on the edges the sampling mechanism follows.
The consequences are not merely theoretical.
\citet{lee2006statistical} compare node, edge, and link-tracing sampling across several scale-free networks and find that vertices reached by tracing edges outward from an initial sample have disproportionately high degree, systematically biasing estimated degree distributions, path lengths, assortativity, and clustering coefficients.
\citet{handcock2010modeling} formalize when a network sampling design can be disregarded, introducing \emph{amenability} as a network-sampling analogue of the ignorability condition of \citet{rubin1976inference}.
Under an amenable design, likelihood-based inference can proceed from the observed data alone, with the likelihood expressed as a sum over every possible completion of the unsampled portion of the network.
In general this sum ranges over a combinatorially large space and cannot be evaluated directly.
\citet{handcock2010modeling} approximate it by Markov Chain Monte Carlo, and beyond a handful of special cases, deriving it in closed form remains open.

This sum admits a closed form, however, in at least one important case.
\citet[Research Problem 3.1]{crane2018network} poses as an open problem the derivation of the distribution induced by $r$-step snowball sampling applied to a randomly chosen vertex of an Erd\H{o}s--R\'{e}nyi population, proposing it as a tractable starting point before richer population models.
In snowball sampling, an initial set of seed vertices is enquired for their neighbours; those neighbours are added to the sample and enquired for their own neighbours; and this process repeats for a fixed number of waves \citep{goodman1961snowball}.
Depending on the branching rule, a sampled vertex's neighbours may be recruited in full, up to a fixed number, or each with some probability \citep{illenberger2012estimating, oguzalper2023snowball, handcock2011concept, heckathorn2011snowball}.
Full recruitment, in which every identified neighbour is added to the sample, is the design used by most of the model-based network sampling literature \citep{illenberger2012estimating,frank2014survey}, including \citet{crane2018network}.

We consider full-neighbourhood recruitment, the setting in which this population-level sum admits an exact, tractable enumeration and a much simpler description than the sum itself might suggest.
We adopt the Erd\H{o}s--R\'enyi model as a tractable starting point: it is unrealistic for most real-world networks, but it yields a likelihood, an estimator, and confidence intervals in closed form.
Under this model, the edge probability $\pi$ is exactly the expected global density of the population network:
It defines a curved exponential family in the edge probability $\pi$, with the number of observed edges and the number of vertex pairs whose adjacency the design reveals together forming a low-dimensional minimal sufficient statistic.
On this basis, we derive the maximum likelihood estimator of $\pi$ that correctly accounts for the sampling design, which we propose as a design-corrected estimator of the population density.
The snowball-corrected estimator depends on the data only through this minimal sufficient statistic and accounts for the exclusion of the unsampled vertices and the dependence of the sample size on $\pi$, unlike the naive estimator, which treats the sampled network as though it were itself a complete Erd\H{o}s--R\'{e}nyi graph and hence estimates the density of the sample rather than of the population.
A simulation study presented in section \ref{subsec:er_simulation_verify} shows that this correction substantially eliminates the bias relative to the naive estimator.

Model-based estimation from snowball samples has also been studied for exponential random graph models, a richer class permitting dyad dependence, using either full-likelihood approximations by Markov chain Monte Carlo \citep{handcock2010modeling} or conditional estimators that avoid modelling the sampling mechanism and the population size \citep{pattison2013conditional}.
By contrast, our Erd\H{o}s--R\'{e}nyi setting allows the design, as well as the information carried by the excluded vertices, to be modelled exactly.

A point estimate alone, however, does not convey its uncertainty, and the same design-induced complications that bias the naive estimator also invalidate standard asymptotic confidence intervals built around it.
We therefore construct valid confidence intervals for $\pi$ by inverting a test based on the exact sampling distribution, using Monte Carlo methods for computational tractability at realistic sample sizes.
Simulations confirm that these intervals attain nominal coverage to within Monte Carlo error across a range of edge probabilities and numbers of waves, while being substantially narrower than a naive bracket based on the same test.

A central caveat, which we emphasize throughout, is that the identification of $\pi$ with the global density is specific to the Erd\H{o}s-Rényi assumption. 
If the population network follows a different model, for instance one with heterogeneous degrees or prominent clustering, the parameter targeted by the snowball-adjusted estimator no longer coincides with the population density. 
We demonstrate this lack of robustness in Section~\ref{sec:robustness}, using synthetic networks that depart from the Erd\H{o}s--R\'{e}nyi assumption, as well as an application to a real Facebook ego network.
Notably, both the empirical application and the simulation results show that, even though the snowball-corrected estimator remains biased, it still vastly improves a naive estimator where the full population is identified with the available sample. 
The method should therefore be read as an exact solution under a fully specified model, as well as a benchmark against which the quality of that specification can be assessed, rather than as a model-free density estimator.

The remainder of the paper is organized as follows.
Section~\ref{sec:notation} defines notation and the snowball sampling scheme.
Section~\ref{sec:erdos} derives the exact likelihood of a snowball sample under the Erd\H{o}s--R\'{e}nyi model, shows that it belongs to a curved exponential family with a low-dimensional minimal sufficient statistic, and verifies the derivation via simulation.
Section~\ref{sec:MLE} derives the snowball-corrected maximum likelihood estimator and evaluates its finite-sample performance against the naive estimator.
Section~\ref{sec:mc_ci} develops Monte Carlo confidence intervals for $\pi$ via test inversion and evaluates their empirical coverage.
Section~\ref{sec:robustness} assesses the robustness of the snowball-corrected estimator to misspecification of the population model, using synthetic networks and an application to a real Facebook ego network.
Section~\ref{sec:discussion} concludes with limitations and directions for future work.

\section{Notation and Sampling Scheme}
\label{sec:notation}

We consider a population network represented by an undirected graph $G = (V, E)$, where $V = \{1, \dots, N\}$ is the vertex set and $E$ is the edge set.
For any finite set $S$, we write
\begin{equation*}
    S^2 = \{\{i,j\} : i,j \in S,\ i \neq j\}
\end{equation*}
for the set of unordered pairs of distinct elements of $S$.
The edge set satisfies $E \subseteq V^2$, and contains no self-loops or multiple edges between the same pair of vertices.
The adjacency matrix $Y = (Y_{i,j})_{1 \le i,j \le N}$ is defined by
\begin{equation*}
    Y_{i,j} = Y_{j,i} =
    \begin{cases}
    1 & \text{if} \quad \{i,j\} \in E, \\
    0 & \text{otherwise}.
    \end{cases}
\end{equation*}
We treat the population network as a realization of a random graph, so $Y$ is a random adjacency matrix with observed realization $y$.
An $r$-wave snowball sample from $G$ is constructed recursively. The initial wave
\begin{equation} \label{eq:wave0}
    \V^{(0)} = \{v_0\}
\end{equation}
consists of a single vertex $v_0 \in V$ (the \textit{ego}), which is either drawn at random from $V$ according to some sampling scheme, or fixed in advance.
When $v_0$ is drawn at random, all subsequent inference is conducted conditionally on the observed $v_0$. We assume throughout the paper that this selection is exogenous, i.e., independent of the network structure.
The first wave comprises all neighbors of the ego,
\begin{equation} \label{eq:wave1}
    \V^{(1)} = \{j \in V : \{v_0, j\} \in E\},
\end{equation}
and, for $k = 2, \dots, r$, the $k$-th wave includes all neighbors of vertices in $\V^{(k-1)}$ not already present in earlier waves,
\begin{equation} \label{eq:wavek}
    \V^{(k)} = \Bigg( \bigcup_{i \in \V^{(k-1)}} \{ j \in V : \{i,j\} \in E \} \Bigg) \setminus \Bigg( \bigcup_{s=0}^{k-1} \V^{(s)} \Bigg).
\end{equation}
Along with \eqref{eq:wave0}--\eqref{eq:wavek}, we define the sampled graph as $G^{(r)} = \left(\V^{(\le r)}, E^{(r)}\right)$, where
\begin{equation*}
    \V^{(\le r)} = \bigcup_{k=0}^r \V^{(k)}, \quad \text{and} \quad E^{(r)} \subseteq \bigl(\bigcup_{k=0}^r \V^{(k)}\bigr)^2.
\end{equation*}
Finally, the sampled adjacency matrix is defined as
\begin{equation} \label{eq:sampledAdjacency}
    Y^{(r)}_{i,j} = Y^{(r)}_{j,i} =
    \begin{cases}
        1 & \text{if} \quad i,j \in \V^{(\le r)}, \ \{i,j\} \in E^{(r)}, \\
        0 & \text{if} \quad i,j \in \V^{(\le r)}, \ \{i,j\} \notin E^{(r)}.
    \end{cases}
\end{equation}
Note that the wave sets $\V^{(1)}, \dots, \V^{(r)}$ are themselves random, as their composition depends on the unknown parameter of the population network, and it is the joint randomness of $\rY$ and the wave sets which poses non-trivial challenges for statistical inference from snowball samples.

\begin{figure}[!ht]
    \centering
    \includegraphics[scale = 0.45]{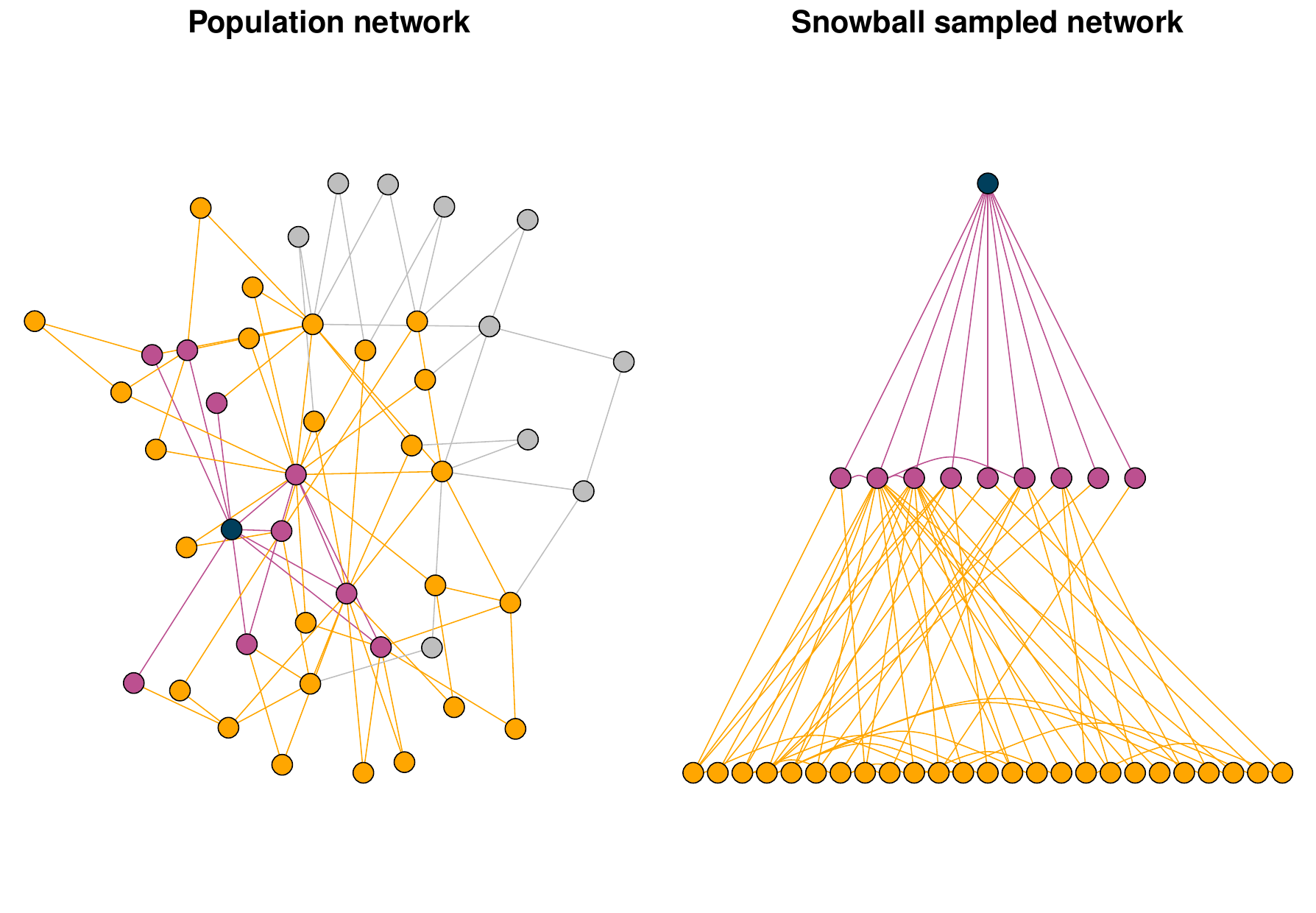}
    \caption{Graphical illustration of a 2-wave snowball sample.}
    \label{fig:pa_graph}
\end{figure}

Figure~\ref{fig:pa_graph} illustrates a 2-wave snowball sample from a population network, with vertices color-coded by wave membership.

\section{Probability of a Snowball Sample From an Erd\H{o}s--R\'enyi Network}
\label{sec:erdos}

We assume that the population network follows an Erd\H{o}s--R\'enyi (ER) model with $N$ vertices.
Under the ER model, each unordered vertex pair $\{i,j\} \in V^2$ is connected by an edge with probability $\pi$, so that the edge indicators $Y_{i,j}$ are independent and identically distributed Bernoulli random variables:
\begin{equation} \label{eq:ERmodel}
    Y_{i,j} \sim \text{Bernoulli}(\pi) \quad \text{for all} \quad \{i,j\} \in V^2,
\end{equation}
where $0 < \pi < 1$ is the unknown edge probability.

Suppose an $r$-wave snowball sample is drawn from an ER population network with edge probability $\pi$ as in \eqref{eq:ERmodel}.
Our objective is to estimate $\pi$ using only the information contained in the sampled subgraph.
A naive approach would treat the snowball sample as representative of the entire population and estimate $\pi$ by the observed density of the sampled subgraph. However, this approach is typically biased, particularly when the true density is low or when the snowball sample covers only a small fraction of the population network. The bias arises because snowball sampling over-represents vertices with high degree, inflating the observed edge density relative to the true population value $\pi$.
As an illustration, the population network in Figure~\ref{fig:pa_graph} comprises 45 vertices and 87 edges, giving a true density of $\pi \approx 0.09$, whereas the naive estimator based on the snowball sample yields $\hat\pi_{\text{naive}} \approx 0.11$.

To account for the sampling bias, we consider the joint probability of the wave sets $\V^{(1)}, \dots, \V^{(r)}$ and the induced adjacency matrix $Y^{(r)}$ conditional on the ego set $\V^{(0)}$ as defined in \eqref{eq:wave0}--\eqref{eq:sampledAdjacency}.
The likelihood of $\pi$ given the snowball sample is
\begin{equation} \label{eq:likelihood-def}
    \scalebox{0.90}{$L(\pi) = P\big(Y^{(r)} = y^{(r)},\V^{(1)} = V^{(1)}, \dots, \V^{(r)} = V^{(r)} \mid \V^{(0)} = V^{(0)},\ \pi\big),$}
\end{equation}
where $y^{(r)}, V^{(0)}, \dots, V^{(r)}$ are the observed realizations of $Y^{(r)}, \V^{(0)}, \dots, \V^{(r)}$ respectively.

For notational convenience, we suppress the dependence on $\pi$ and write probabilities as functions of observed realizations only, e.g., $P(y^{(r)}, V^{(1)}, \dots, V^{(r)} \mid V^{(0)})$ in place of the right-hand side of \eqref{eq:likelihood-def}.

Factorizing the joint probability yields
\begin{equation} \label{eq:factorization}
    P\big(y^{(r)}, V^{(1)}, \dots, V^{(r)} \mid V^{(0)}\big) = P\big(y^{(r)} \mid V^{(0)}, \dots, V^{(r)}\big) P\big(V^{(1)}, \dots, V^{(r)} \mid V^{(0)}\big),
\end{equation}
where the second factor on the right-hand side of \eqref{eq:factorization} is the marginal probability that the wave sets $\V^{(1)}, \dots, \V^{(r)}$ equal $V^{(1)}, \dots, V^{(r)}$.
Notably, this term also depends on $\pi$ and is the source of bias in the naive approach, which ignores it.

\subsection{Marginal Probability of Wave Sets}
\label{sec:marginal-waves}

The marginal probability of the wave sets in \eqref{eq:factorization} can be written as a product of conditional probabilities:
\begin{equation} \label{eq:wave-chain}
    P\big(V^{(1)}, \dots, V^{(r)} \mid V^{(0)}\big) = P\big(V^{(1)} \mid V^{(0)}\big) \prod_{k=2}^r P\big(V^{(k)} \mid V^{(0)}, \dots, V^{(k-1)}\big).
\end{equation}
We now address each term in \eqref{eq:wave-chain} separately.

Given the ego wave $\V^{(0)} = V^{(0)}$, for any vertex $i \in V \setminus V^{(0)}$ to be included in $V^{(1)}$, there must be an edge from $i$ to the ego, which under \eqref{eq:ERmodel} occurs with probability $\pi$ independently across all vertices in $V \setminus V^{(0)}$.
In the terminology of survey sampling, the distribution of $V^{(1)}$ given $V^{(0)}$ corresponds to a Bernoulli sampling design with success probability $\pi$ over $V \setminus V^{(0)}$ \citep{arnab2017survey}.

Define the size of wave $k$ as the cardinality of the observed wave $V^{(k)}$:
\begin{equation*}
    n_k = |V^{(k)}|, \qquad k = 0, \dots, r,
\end{equation*}
with $n_0 = 1$ by \eqref{eq:wave0}.
Then, the probability of observing $\V^{(1)} = V^{(1)}$ given $\V^{(0)} = V^{(0)}$ is
\begin{align*}
    P(V^{(1)} \mid V^{(0)}) &= \prod_{i \in V^{(1)}} P(i \ \text{has an edge to} \ V^{(0)}) \prod_{j \in V \setminus (V^{(0)} \cup V^{(1)})} P(j \ \text{has no edge to} \ V^{(0)}) \\
    &= \prod_{i \in V^{(1)}} \pi \prod_{j \in V \setminus (V^{(0)} \cup V^{(1)})} (1 - \pi) = \pi^{n_1} (1 - \pi)^{N - n_1 - 1} \numberthis \label{eq:P-wave1}
\end{align*}

Consider now the conditional probability $P(V^{(k)} \mid V^{(0)}, \dots, V^{(k-1)})$ for $k = 2, \dots, r$.
Unlike the case of $\V^{(1)} \mid \V^{(0)} = V^{(0)}$, a vertex $i \in \V^{(k)}$ may have multiple edges leading to $V^{(k-1)}$, since $V^{(k-1)}$ can consist of several vertices.
To be included in $\V^{(k)}$, vertex $i$ must have at least one edge to $V^{(k-1)}$.
Given $V^{(0)}, \dots, V^{(k-1)}$, the distribution of $\V^{(k)}$ again corresponds to a Bernoulli sampling design over $V \setminus \bigl(\bigcup_{s=0}^{k-1} V^{(s)}\bigr)$, with success probability equal to the probability that vertex $i$ has at least one edge to $V^{(k-1)}$:
\begin{align*}
    P(V^{(k)} \mid V^{(0)}, \dots, V^{(k-1)}) &= \prod_{i \in V^{(k)}} P(i \text{ has at least one edge to } V^{(k-1)}) \\
    &\quad \times \prod_{j \in V \setminus (\bigcup_{s = 0}^{k} V^{(s)})} P(j \text{ has no edges to } V^{(k-1)}) \\
    &= \prod_{i \in V^{(k)}} [1 - \prod_{t \in V^{(k-1)}} (1 - \pi)] \prod_{j \in V \setminus \bigcup_{s = 0}^{k} V^{(s)}} \prod_{t \in V^{(k-1)}} (1 - \pi) \\
    &= [1 - (1 - \pi)^{n_{k-1}}]^{n_{k}} (1 - \pi)^{n_{k - 1}\left(N - \sum_{s = 0}^{k} n_s\right)}. \numberthis
    \label{eq:P-wavek}
\end{align*}
Substituting \eqref{eq:P-wave1} and \eqref{eq:P-wavek} into \eqref{eq:wave-chain} gives
\begin{align*}
    P\big(V^{(1)}, \dots, V^{(r)} \mid V^{(0)}\big) &= \pi^{n_1} (1 - \pi)^{N - n_1 - 1} \prod_{k=2}^r \big[1 - (1-\pi)^{n_{k-1}}\big]^{n_k} (1-\pi)^{n_{k-1} \left( N - \sum_{s=0}^k n_s \right)} \\
    &= \prod_{k=1}^r \big[1 - (1-\pi)^{n_{k-1}}\big]^{n_k} (1-\pi)^{n_{k-1} \left( N - \sum_{s=0}^k n_s \right)} \numberthis \label{eq:wave_marginal}
\end{align*}
where the $k=1$ term recovers \eqref{eq:P-wave1} because $(1-\pi)^{N-n_1-1} = (1-\pi)^{n_0[N-(n_1+n_0)]}$ and $\pi^{n_1} = [1-(1-\pi)^{n_0}]^{n_1}$ since $n_0 = 1$ by \eqref{eq:wave0}.

\subsection{Conditional Probability of the Sampled Adjacency Matrix}
\label{subsec:cond_prob}

The realized wave sets $V^{(0)}, \dots, V^{(r)}$ together encompass all vertices of the sampled graph as a realization of $\V^{(\le r)}$:
\begin{equation*}
    V^{(\le r)} = \bigcup_{k = 0}^r V^{(k)}
\end{equation*}
As such, they also determine the support of the conditional distribution of $\rY$.
In particular, knowing which vertices belong to which wave allows us to determine the possible configurations of the adjacency matrix $\rY$ under the snowball sampling mechanism.
These constraints follow directly from the definition of snowball sampling in Section~\ref{sec:notation}:
\begin{enumerate}
    \item \textsl{No edges between non-adjacent waves:} for any $i \in V^{(k)}$ and $j \in V^{(s)}$ with $|k - s| \geq 2$, we have $\rY_{i,j} = 0$.\label{constraint1}
    \item \textsl{At least one edge to the preceding wave:} for any $i \in V^{(k)}$ with $k \in \{1, \dots, r\}$, there exists $j \in V^{(k-1)}$ such that $\rY_{i,j} = 1$.\label{constraint2}
\end{enumerate}
These constraints imply that the conditional distribution of $\rY$ given the wave sets differs from the standard ER model due to its restricted support.
Since the constraints apply to different subsets of vertex pairs, we define the following partition of $\left(V^{(\le r)}\right)^2$ into three disjoint sets:
\begin{itemize}
    \item \textsl{Vertex pairs within the same wave:}
    \begin{equation}\label{eq:defW}
        W = \left\{\{i,j\} \in \left(V^{(\le r)}\right)^2 : i, j \in V^{(t)},\ t = 1, \dots, r \right\}
    \end{equation}
    \item \textsl{Vertex pairs between adjacent waves:}
    \begin{equation}\label{eq:defA}
        A = \left\{\{i,j\} \in \left(V^{(\le r)}\right)^2 : i \in V^{(t)},\ j \in V^{(t-1)},\ t = 1, \dots, r \right\}
    \end{equation}
    \item \textsl{Vertex pairs between non-adjacent waves:}
    \begin{equation}
    \label{eq:defJ}
        J = \left(V^{(\le r)}\right)^2\setminus (W \cup A)
    \end{equation}
\end{itemize}

For any subset of the vertex pairs $S \subseteq \left(V^{(\le r)}\right)^2$, we write $\rY_S$ for the entries of $\rY$ indexed by $S$, and $\ry_S$ for its realization.
For notational simplicity, we maintain the convention of omitting random variables when writing probabilities, so that, for example, $P(\rY = \ry)$ is written as $P(\ry)$.

Using \eqref{eq:defW}--\eqref{eq:defJ}, the conditional probability of $Y^{(r)} = y^{(r)}$
given the wave sets factors as
\begin{align*}
    P\big(y^{(r)} \mid V^{(0)}, \dots, V^{(r)}\big) &= P\big(y^{(r)}_A \mid y^{(r)}_W, y^{(r)}_J, V^{(0)}, \dots, V^{(r)}\big) \\
    &\quad \times P\big(y^{(r)}_W \mid y^{(r)}_J, V^{(0)}, \dots, V^{(r)}\big) \\
    &\quad \times P\big(y^{(r)}_J \mid V^{(0)}, \dots, V^{(r)}\big).
\end{align*}
The constraints of the snowball sampling design imply that $\rY_J = \ry_J = \boldsymbol{0}$ with probability one, so that the above expression simplifies to
\begin{align*}
    P(\ry \mid V^{(0)}, \dots, V^{(r)}) &= P(\ry_A \mid \ry_W, \ry_J = \boldsymbol{0}, V^{(0)} \dots, V^{(r)}) \\
    &\quad \times P(\ry_W \mid \ry_J = \boldsymbol{0}, V^{(0)}, \dots, V^{(r)}), \numberthis \label{eq:er_conditional_prob}
\end{align*}
for any $\ry$ such that $\ry_J = \boldsymbol{0}$, and zero otherwise.

The probability of $\rY_W = \ry_W$ is independent of the constraints, as the within-wave edges are not subject to any constraints under the snowball sampling design.
Under the ER model, all edges are independent Bernoulli variables, so that
\begin{equation} \label{eq:condprobW}
    P(\ry_W \mid \ry_J = \boldsymbol{0}, V^{(0)}, \dots, V^{(r)}) = P(\ry_W) = \prod_{\{i,j\} \in W} \pi^{\ry_{i,j}}(1 - \pi)^{1 - \ry_{i,j}}.
\end{equation}

Similarly, the edge indicators in $\rY_A$ are independent of $\rY_J$ and $\rY_W$ given the wave sets, so the first term on the right-hand side of \eqref{eq:er_conditional_prob} simplifies to
\begin{equation*}
    P(\ry_A \mid \ry_W, \ry_J = \boldsymbol{0}, V^{(0)} \dots, V^{(r)}) = P(\ry_A \mid V^{(0)} \dots, V^{(r)}).
\end{equation*}

Unlike the conditional probability of $\rY_W$, however, $P(\ry_A \mid V^{(0)}, \dots, V^{(r)})$ is not simply the product of the probabilities over each vertex pair in $A$, as the snowball sampling design imposes constraints on $\rY_A$ that induce dependencies between edges in $A$: every vertex in wave $k$ must connect to at least one vertex in wave $k-1$.

To handle this, let $v(i)\in \{0, \dots, r\}$ be the wave index of vertex $i$, and define
\begin{equation}\label{eq:defAi}
    A(i) = \{\{i,j\} \in A : j \in V^{(v(i) - 1)}\}
\end{equation}
as the set of vertex pairs in $A$ that include vertex $i$ and all vertices in the preceding wave.
Then $A$ can be partitioned into the union of disjoint subsets:
\begin{equation*}
    A = \bigcup_{i \in V^{(\le r)}} A(i).
\end{equation*}

The constraints \ref{constraint1}--\ref{constraint2} of the snowball sampling design imply that $\sum_{\{u,l\} \in A(i)} \rY_{u,l} > 0$ for every $i \in V^{(\le r)}$ with probability one.
Therefore:
\begin{align*}
    P(\ry_{A(i)} \mid V^{(0)}, \dots, V^{(r)}) &= P\Big(\ry_{A(i)} \mid \sum_{\{u,l\} \in A(i)} \rY_{u,l} > 0\Big) = \frac{P(\ry_{A(i)})}{1 - P(\rY_{A(i)} = \boldsymbol{0})} \\
    &= \frac{\prod_{\{u,l\} \in A(i)} \pi^{\ry_{u,l}}(1 - \pi)^{1 - \ry_{u,l}}}{1 - (1 - \pi)^{n_{v(i) - 1}}},
\end{align*}
where $n_{v(i) - 1} = |V^{(v(i) - 1)}|$ is the number of vertices in the preceding wave of vertex $i$.

Since the sets $A(i)$ are disjoint and the constraints apply to each $A(i)$ separately, the conditional probability of $\rY_A = \ry_A$ given the wave sets is
\begin{align*}
    P(\ry_A \mid V^{(0)}, \dots, V^{(r)}) &= \prod_{i \in \bigcup_{k = 1}^r V^{(k)}} \frac{\prod_{\{u,l\} \in A(i)} \pi^{\ry_{u,l}}(1 - \pi)^{1 - \ry_{u,l}}}{1 - (1 - \pi)^{n_{v(i) - 1}}} \\
    &= \frac{\prod_{\{i,j\} \in A} \pi^{\ry_{i,j}}(1 - \pi)^{1 - \ry_{i,j}}}{\prod_{k = 1}^r [1 - (1 - \pi)^{n_{k - 1}}]^{n_k}}, \numberthis \label{eq:condprobA}
\end{align*}
where the denominator of the last equality follows from the fact that each of the $n_k$ vertices in wave $k$ contributes the same term $1 - (1 - \pi)^{n_{k-1}}$.

Combining \eqref{eq:condprobA} and \eqref{eq:condprobW} in \eqref{eq:er_conditional_prob}, we obtain
\begin{equation} \label{eq:conditional_mat}
    P(\ry \mid V^{(0)}, \dots, V^{(r)}) =  \frac{\prod_{\{i,j\} \in A \cup W} \pi^{\ry_{i,j}}(1 - \pi)^{1 - \ry_{i,j}}}{\prod_{k = 1}^r [1 - (1 - \pi)^{n_{k - 1}}]^{n_k}},
\end{equation}
which equals zero if $\ry$ violates any of the constraints \ref{constraint1} and \ref{constraint2}.

\subsection{Probability of a Snowball Sample}
\label{subsec:prob_snow}
Following the factorization in \eqref{eq:factorization}, the joint probability of the sampled adjacency matrix and the wave sets can be expressed as a product of \eqref{eq:conditional_mat} and \eqref{eq:wave_marginal}:
\begin{align*}
    P(\ry, V^{(1)}, \dots, V^{(r)} \mid V^{(0)}) &= \frac{\prod_{\{i,j\} \in W \cup A} \pi^{\ry_{i,j}}(1 - \pi)^{1 - \ry_{i,j}}}{\prod_{k = 1}^r [1 - (1 - \pi)^{n_{k - 1}}]^{n_k}} \\
    &\quad \times \prod_{k = 1}^r [1 - (1 - \pi)^{n_{k - 1}}]^{n_k} (1 - \pi)^{n_{k - 1}(N - \sum_{s = 0}^{k} n_s)}\numberthis
    \label{eq:jointprob}
\end{align*}
The expression can be considerably simplified by using the number of unsampled units
\begin{equation*}
    U = |V \setminus V^{(\le r)}| = N - \sum_{k = 0}^r n_k.
\end{equation*}
The exponent on the right-hand side of \eqref{eq:jointprob} can thus be written as:
\begin{align*}
    \sum_{k = 1}^r n_{k - 1}(N - \sum_{s = 0}^k n_k) &= \sum_{k = 1}^r \left[n_{k - 1}(U + \sum_{s = 0}^r n_s - \sum_{l = 0}^k n_l)\right] \\
    &= n_{r-1} U  + \sum_{k = 1}^{r - 1} \left[n_{k - 1}U + n_{k - 1}\sum_{s = k + 1}^r n_s\right] \\
    &= U \sum_{k = 1}^r n_{k - 1} + \sum_{k = 1}^{r - 1} \left[n_{k - 1} \sum_{s = k + 1}^r n_s\right]
    \numberthis
    \label{eq:sum_of_nk}
\end{align*}
The second sum is the cardinality of the set of vertex pairs between non-adjacent waves $J$ defined in \eqref{eq:defJ}.
This can be shown directly by enumerating the number of vertices from the non-adjacent waves for each wave. For the ego set $V^{(0)}$, there are $\sum_{s = 2}^r n_s$ vertices from the non-adjacent waves.
Similarly, for each of the $n_1$ vertices in the first wave $V^{(1)}$, there are $\sum_{s = 3}^r n_s$ vertices from the non-adjacent waves.
Extending this calculation to all $r - 2$ waves yields
\begin{equation} \label{eq:cardJ}
    |J| = \sum_{s = 2}^r n_s + n_1 \sum_{s = 3}^r n_s + \dots + n_{r-2} n_r = \sum_{k = 0}^{r - 2} n_k \sum_{s = k + 2}^r n_s = \sum_{k = 1}^{r - 1} n_{k - 1} \sum_{s = k + 1}^r n_s.
\end{equation}
Putting together \eqref{eq:sum_of_nk} with \eqref{eq:cardJ} and substituting into \eqref{eq:jointprob}, the joint probability of the wave sets and the sampled adjacency matrix can be written as
\begin{equation*}
    P(\ry, V^{(1)}, \dots, V^{(r)} \mid V^{(0)}) = \left[\prod_{\{i,j\} \in W \cup A} \pi^{\ry_{i,j}}(1 - \pi)^{1 - \ry_{i,j}}\right] (1 - \pi)^{U \sum_{k = 0}^{r - 1} n_k} (1 - \pi)^{|J|}
\end{equation*}
Finally, because $y_{i,j} = 0$ for all $\{i,j\} \in J$ by definition, we can incorporate the right-most term into the product as
\begin{equation} \label{eq:er_joint_prob_complete}
    P(\ry, V^{(1)}, \dots, V^{(r)} \mid V^{(0)}) = \left[\prod_{\{i,j\} \in W \cup A \cup J} \pi^{\ry_{i,j}}(1 - \pi)^{1 - \ry_{i,j}}\right] (1 - \pi)^{U \sum_{k = 0}^{r - 1} n_k}.
\end{equation}

The probability in \eqref{eq:er_joint_prob_complete} admits a direct interpretation.
The product term is the joint probability of edges and non-edges in the observed subgraph under the Erd\H{o}s-R\'enyi population model \eqref{eq:ERmodel}, while the term $(1 - \pi)^{U \sum_{k = 0}^{r - 1} n_k} = [(1 - \pi)^{\sum_{k = 0}^{r - 1} n_k}]^U$ is the joint probability of non-edges between the unsampled nodes and the vertices sampled in waves $0$ through $r - 1$.
This term encodes the fact that the unsampled vertices were excluded from the sample, since inclusion would require an edge to some vertex in $V^{(0)}, \dots, V^{(r - 1)}$.
Thus the joint probability \eqref{eq:er_joint_prob_complete} captures all information available from the sample: the observed subgraph together with the exclusion of the $U$ unsampled nodes.

\subsection{Joint Distribution as a Curved Exponential Family}\label{subsec:curved}
It can be shown that the right-hand side of \eqref{eq:er_joint_prob_complete} depends on the data only through two scalar quantities.
Writing  $n = (n_1, \dots, n_r)$, we define the total number of constrained vertex pairs as
\begin{equation}
\label{eq:T}
    T(n) = |W| + |A| + |J| + U \sum_{k = 0}^{r-1} n_k,
\end{equation}
Then, \eqref{eq:er_joint_prob_complete} can be written compactly as
\begin{align*}
    P(\ry, V^{(1)}, \dots, V^{(r)} \mid V^{(0)}) &= \pi^{|E^{(r)}|} (1 - \pi)^{T(n) - |E^{(r)}|} \ \mathds{1}\{\ry \in \mathcal{S}(n)\}\\
    &= \exp\left\{ |E^{(r)}| \log\left(\frac{\pi}{1 - \pi}\right) + T(n) \log(1-\pi) \right\} \\
    &\quad \times \mathds{1}\{\ry \in \mathcal{S}(n)\} \numberthis \label{eq:exp_family_form}
\end{align*}
where $\mathcal{S}(n)$ is the set of adjacency matrices consistent with the constraints \ref{constraint1} and \ref{constraint2}.

The probability mass function \eqref{eq:exp_family_form} is of a two-parameter exponential family with natural statistic $(|E^{(r)}|, T(n))$ and natural parameters $\log \frac{\pi}{1-\pi}$ and $\log(1-\pi)$.
Because the dimension of the natural statistic exceeds the dimension of the parameter $\pi$, the distribution belongs to a curved exponential family \citep{efron1975defining}, which distinguishes it from the ordinary Erd\H{o}s-R\'enyi model on a fixed vertex set.
The curvature arises because the sample size itself is informative about $\pi$: larger $\pi$ tends to produce larger waves and hence larger $T(n)$.

By the Fisher--Neyman factorization theorem, since \eqref{eq:exp_family_form} depends on the data only through $|E^{(r)}|$ and $T(n)$, with the remaining factor $\mathds{1}\{\ry \in \mathcal{S}(n)\}$ not depending on $\pi$, the pair $(|E^{(r)}|, T(n))$ is sufficient for $\pi$.
It is also minimal: the support of the family is the same for every $\pi \in (0,1)$, so it suffices to compare the likelihood ratio for $\ry \in \mathcal{S}(n)$ and $y^{(r)^\prime} \in \mathcal{S}(n^\prime)$, where it reduces to
\begin{equation*}
    \left(\frac{\pi}{1-\pi}\right)^{|E^{(r)}| - |E^{(r)}|'} (1-\pi)^{T(n) - T(n)'}.
\end{equation*}
This is constant in $\pi$ for all $\pi \in (0,1)$ if and only if $|E^{(r)}| = |E^{(r)}|'$ and $T(n) = T(n)'$, because $\log\frac{\pi}{1-\pi}$ and $\log(1-\pi)$ are not affinely related as functions of $\pi$, i.e. there is no non-trivial linear combination of the two that is constant on $(0,1)$.
By the standard characterization of minimal sufficiency for exponential families, $(|E^{(r)}|, T(n))$ is therefore minimal sufficient for $\pi$ \citep{casella2002principles}.

Two remarks are worth making explicit.
First, although the individual wave sizes $n_1, \dots, n_r$ are part of the raw data and are trivially sufficient, they are not minimal: two wave-size vectors $n \neq n^\prime$ with $T(n) = T(n^\prime)$ carry the same information about $\pi$, even though the support of the sampled adjacency matrices $\rY$, given the wave sizes, may differ.
Second, as later shown in Section~\ref{sec:MLE}, the maximum likelihood estimator $\hat{\pi}$ is a function of the minimal sufficient statistic alone, as it must be for any exponential family.

\subsection{Simulation-Based Verification}
\label{subsec:er_simulation_verify}

The explicit forms of the chain probabilities of the wave sets \eqref{eq:wave_marginal} and the conditional distribution of the sampled adjacency matrix \eqref{eq:conditional_mat} suggest a straightforward algorithm for directly simulating snowball samples from an ER population without generating the full population graph.

The wave sets $\V^{(0)}, \dots, \V^{(r)}$ can be generated as follows:
\begin{enumerate}
    \item Choose an ego vertex $i_0$ from the population vertex set $V$ independently of the network structure and set $V^{(0)} = \{i_0\}$.
    \item Draw the first wave $\V^{(1)}$ by vertices $i \in V \setminus V^{(0)}$ with probability $\pi$.
    \item For each $k = 2, \dots, r$:
    \begin{enumerate}
        \item[3.1] Draw the wave $\V^{(k)}$ by including vertices $i \in V \setminus \bigcup_{s = 0}^{k - 1} V^{(s)}$ with probability $1 - (1 - \pi)^{n_{k - 1}}$, where $n_{k - 1} = |V^{(k-1)}|$.
    \end{enumerate}
\end{enumerate}
Once the wave sets are generated, simulate the adjacency matrix $\rY$ as follows:
\begin{enumerate}
    \item For each $\{i,j\}$ in $J$, set $\rY_{i,j} = 0$.
    \item For each $\{i,j\}$ in $W$, draw $\rY_{i,j}$ from a Bernoulli distribution with parameter $\pi$.
    \item For each $i \in \bigcup_{k = 1}^r V^{(k)}$:
    \begin{enumerate}
        \item[3.1] Draw the number of edges $m_i$ from a zero-truncated binomial distribution with parameters $n_{v(i) - 1}$ and $\pi$, where $n_{v(i) - 1} = |V^{(v(i) - 1)}|$ is the number of vertices in the preceding wave.
        \item[3.2] Draw $m_i$ vertex pairs $\{i,j\}$ from $A(i)$ (as defined in \eqref{eq:defAi}) uniformly at random without replacement and set $\rY_{i,j} = 1$ for each of them.
    \end{enumerate}
\end{enumerate}
The resulting adjacency matrix $\rY$ is a realization of the $r$-wave snowball sampling design from an ER graph with $N$ vertices and edge probability $\pi$.

To verify the correctness of our derivations, we compare the joint distribution of the minimal sufficient statistic $(|E^{(r)}|, T(n))$ identified in Section~\ref{subsec:curved} under the proposed per-wave simulation algorithm and under direct snowball sampling from a fully generated ER graph.
Direct snowball sampling proceeds as follows:
\begin{enumerate}
    \item Generate an Erd\H{o}s-R\'{e}nyi network with $N$ vertices and edge probability $\pi$.
    \item Choose an ego vertex $i_0$ from the population vertex set $V = \{1, \dots, N\}$ with robability independent of the network structure.
    \item Draw an $r$-wave snowball sample from $i_0$ following the sampling scheme in Section~\ref{sec:notation}.
\end{enumerate}

We set $r = 3$ and generate $M = 10{,}000$ snowball samples from an ER graph with $N = 15{,}000$ vertices and $\pi = 0.001$ under each of the two methods, and test equality of the resulting empirical joint distributions of $(|E^{(r)}|, T(n))$ using a permutation test based on the two-sample energy statistic.

For independent $d$-dimensional random vectors $X, Z$ with distribution functions $F$ and $H$, and their independent and identically distributed copies $X^\prime, Z^\prime$ with $\mathbb{E}[\lVert X\rVert] < \infty$, $\mathbb{E}[\lVert Z\rVert] < \infty$, the energy distance
\begin{equation*}
    \mathcal{E}(F,H) = 2\mathbb{E}[\lVert X-Z\rVert] - \mathbb{E}[\lVert X-X^\prime\rVert] - \mathbb{E}[\lVert Z-Z^\prime\rVert]
\end{equation*}
is nonnegative and equals zero if and only if $X$ and $Z$ are identically distributed.
As such, $\mathcal{E}(F,H)$ provides a characterization of equality of distributions of the multivariate $X$ and $Z$ \citep{szekely2013energy}.
We exploit this here, as a scalar statistic such as a coordinatewise Kolmogorov--Smirnov statistic would not characterize equality of the joint distribution of $(|E^{(r)}|, T(n))$ across the two sampling methods.

Given independent samples $x_1, \dots, x_M$ from $F$ and $z_1, \dots, z_M$ from $H$, the two-sample energy statistic
\begin{equation}
    \mathcal{E}_M(x,z) = \frac{2}{M^2}\sum_{i=1}^m \sum_{j = 1}^M \lVert x_i - z_j\rVert - \frac{1}{M^2}\sum_{i=1}^M \sum_{j = 1}^M \lVert x_i-x_{j}\rVert - \frac{1}{M^2}\sum_{i=1}^M \sum_{j = 1}^M \lVert z_i-z_j\rVert
\end{equation}
is the corresponding plug-in estimator of $\mathcal{E}(F,H)$, and for samples of equal size $M$, the test statistic is
\begin{equation*}
    Q_M(x,z) = \frac{M}{2}\, \mathcal{E}_M(x,z)
\end{equation*}
as defined by \citet{szekely2013energy}, with $H_0 : F = H$ rejected in favor of $H_1 : F \neq H$ for large values of $Q_M$.

In our case, $F$ and $H$ denote the true joint distributions of sufficient statistics $(|E^{(r)}|, T(n))$ under direct snowball sampling and the per-wave simulation algorithm, respectively.
Because the null distribution of $Q_M$ depends on the (unknown) distributions $F$ and $H$ themselves, we approximate the exact $p$-value via permutation \citep{hodges1958significance, schroer1995exact}.
Under $H_0$, the two samples are exchangeable, so we pool the $20{,}000$ observed pairs, repeatedly partition the pooled set into two groups of size $10{,}000$, and recompute the statistic $Q_M^{*}$ on each partition, treating the two groups as if they were the original two samples.
Repeating this process $B$ times yields a collection of statistics $Q_{M,1}^{*}, \dots, Q_{M,B}^{*}$ approximating the null distribution of $Q_M$ under $H_0$.
The Monte Carlo $p$-value is
\begin{equation*}
    \hat{p} = \frac{1}{B}\sum_{b=1}^{B} \mathds{1}\left(Q_{M,b}^{*} \geq Q_M \right),
\end{equation*}
which converges to the exact $p$-value as $B \to \infty$.

\begin{figure}[ht!]
    \centering
    \begin{subfigure}{0.48\linewidth}
        \centering
        \includegraphics[width=\linewidth]{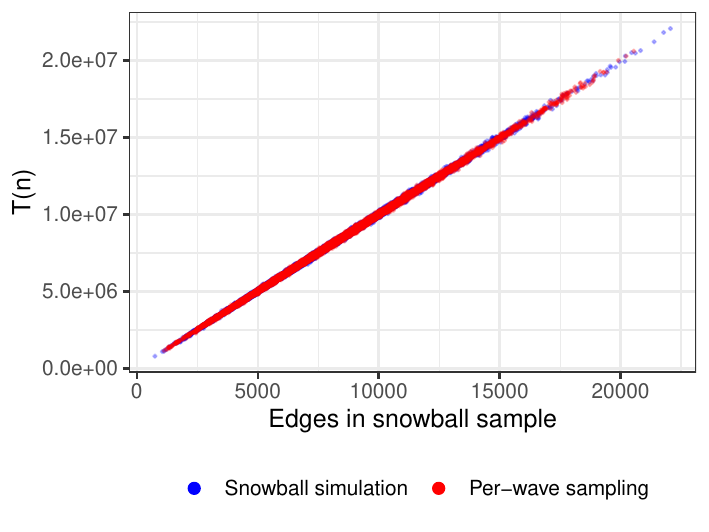}
        \caption{Scatter plot of $(|E^{(r)}|, T(n))$ under direct 3-wave snowball sampling (blue) and the simulation algorithm of Section~\ref{subsec:er_simulation_verify} (red).}
        \label{fig:er_T_scatter}
    \end{subfigure}
    \hfill
    \begin{subfigure}{0.48\linewidth}
        \centering
        \includegraphics[width=\linewidth]{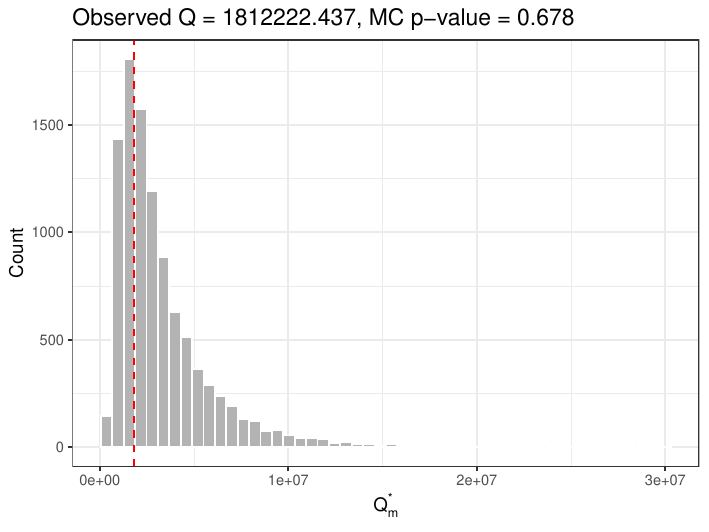}
        \caption{Permutation null distribution of $Q_m^{*}$, with the observed statistic $Q_m$ marked.}
        \label{fig:energy_perm_dist}
    \end{subfigure}
    \caption{Verification of the joint distribution of $(|E^{(r)}|, T(n))$ under direct and per-wave simulated 3-wave snowball sampling, repeated $m = 10{,}000$ times for $\pi = 0.001$ and $N = 15{,}000$. The observed test statistic is $Q_m = 1{,}812{,}222.76$, with Monte Carlo $p$-value $\hat{p} = 0.678$ estimated from $B = 10{,}000$ permutations.}
    \label{fig:er_simulation}
\end{figure}

Figure~\ref{fig:er_simulation} plots the joint scatter of $(|E^{(r)}|, T(n))$ under both simulation methods alongside the permutation null distribution of $Q_M^{*}$ at $B = 10{,}000$, with the observed statistic $Q_M$ marked.
The two point clouds are visually indistinguishable, and the observed statistic $Q_M = 1{,}812{,}222.76$ falls well within the bulk of the permutation null distribution, consistent with a Monte Carlo $p$-value of $\hat{p} = 0.678$.
The null hypothesis of equal joint distributions cannot be rejected at conventional significance levels.

\section{Maximum Likelihood Estimation}
\label{sec:MLE}
Following \eqref{eq:exp_family_form}, the log-likelihood function of $\pi$, up to additive constants, is
\begin{equation} \label{eq:log_like}
    \ell(\pi) = |E^{(r)}| \log \pi + (T(n) - |E^{(r)}|) \log(1 - \pi)
\end{equation}
and the corresponding maximum likelihood estimator can be derived as
\begin{equation} \label{eq:er_snow_mle}
    \hat{\pi} = \frac{|E^{(r)}|}{T(n)} = \frac{\sum_{\{i,j\} \in A \cup W} \ry_{i,j}}{|W| + |A| + |J| + U\sum_{k = 0}^{r - 1} n_k}.
\end{equation}
which is a function of the minimal sufficient statistic $(|E^{(r)}|, T(n))$ alone as noted in Section~\ref{subsec:prob_snow}

For comparison, the naive estimator of $\pi$, which treats the sampled graph as a realization of an ER graph on the vertex set $V^{(\le r)}$, is
\begin{equation} \label{eq:er_naive_mle}
    \hat{\pi}_{\text{naive}} = \frac{\sum_{\{i,j\} \in W \cup A} \ry_{i,j}}{|W| + |A| + |J|}.
\end{equation}

The two estimators are nearly identical in structure with slightly different denominators.
While the naive estimator divides the number of edges by all possible vertex pairs in the observed subgraph, the snowball-corrected estimator \eqref{eq:er_snow_mle} divides by all possible vertex pairs and the absent edges between unsampled nodes and the nodes from waves $0, 1, \dots, r - 1$.
This is exactly the source of the upward bias of the naive estimator, which vanishes as the snowball sample covers more of the population network, i.e., as $U \to 0$.

\begin{figure}[htbp]
\centering
\captionsetup[subfigure]{font=scriptsize}
\begin{subfigure}{\linewidth}
    \centering
    \includegraphics[width = \linewidth]{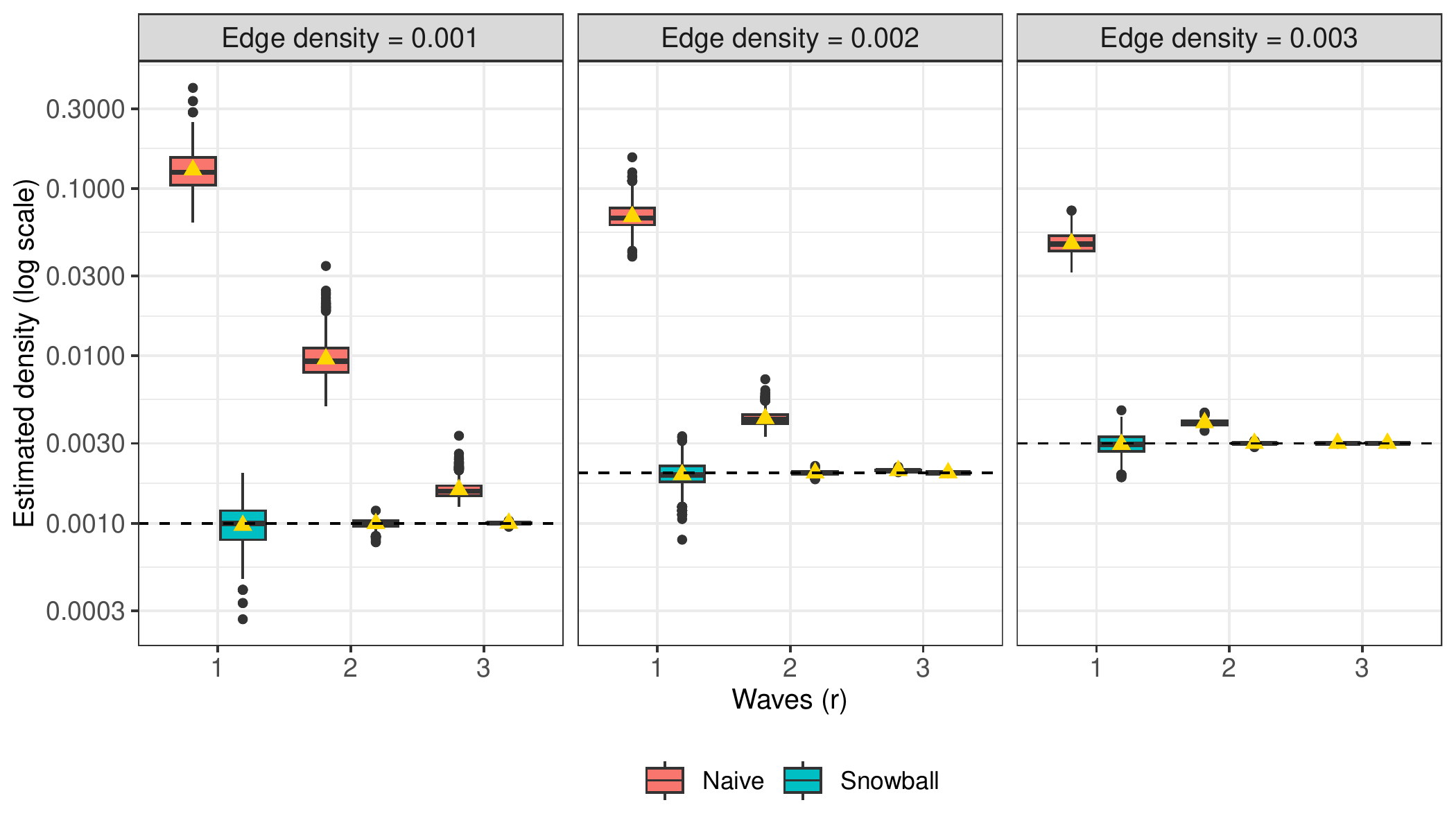}
    \caption{Boxplots of estimates of $\pi$ using the naive \eqref{eq:er_naive_mle} and snowball-corrected \eqref{eq:er_snow_mle} estimators. Dashed lines indicate the true $\pi$.}
    \label{fig:p_boxplot}
\end{subfigure}
\begin{subfigure}{\linewidth}
    \centering
    \includegraphics[scale = 0.3625]{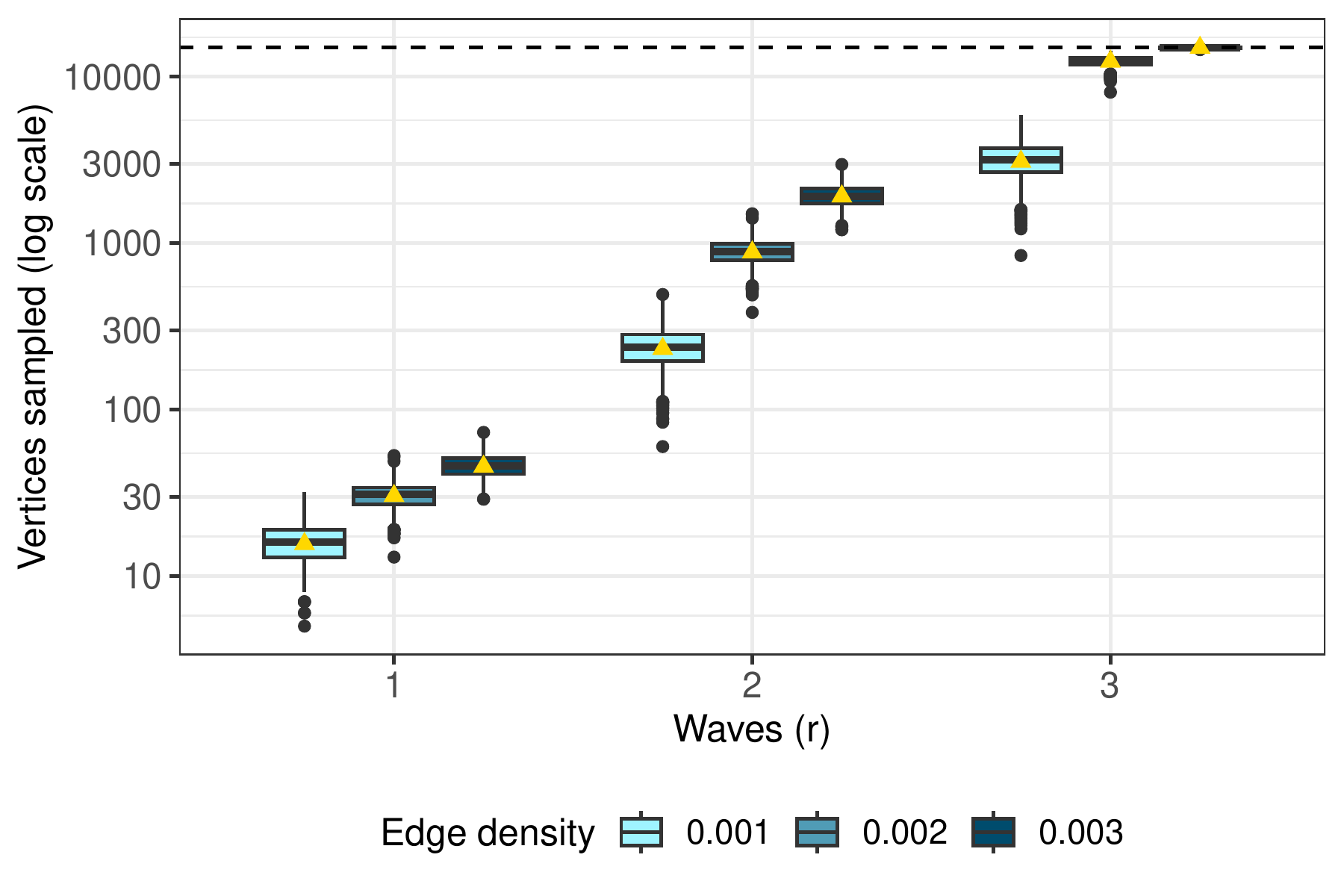}
    \caption{Boxplots of the number of sampled vertices. Dashed line indicates the total number of vertices $N = 15{,}000$.}
    \label{fig:p_nodes}
\end{subfigure}
\caption{Boxplots of the naive and snowball-corrected estimators of $\pi$ and the number of sampled vertices over $1{,}000$ simulations for different values of edge probability $\pi$ and number of waves $r$.
Triangles indicate the means of the distributions.}
\end{figure}

To illustrate the finite-sample behavior of the snowball-corrected estimator \eqref{eq:er_snow_mle} relative to the naive estimator \eqref{eq:er_naive_mle}, we simulate $r$-wave snowball samples from ER graphs with $N$ vertices and edge probability $\pi$.
We set $N = 15{,}000$ and let $r$ vary from $1$ to $3$.
For each $\pi \in \{0.001, 0.002, 0.003\}$ and each $r$, we generate $1{,}000$ ER graphs and draw $r$-wave snowball samples from each.
The values of $\pi$ and $N$ are chosen to reflect the sparsity of real-world networks.
In particular, they ensure that a snowball sample leaves a non-trivial fraction of the population unsampled, so that the difference between the two estimators remains meaningful.
Both estimators are applied to each sample, and the results are summarized in Figure~\ref{fig:p_boxplot}, while Figure~\ref{fig:p_nodes} shows the distribution of the number of sampled vertices for each combination of $\pi$ and $r$.

As expected, the naive estimator overestimates the edge probability $\pi$ unless $r$ or $\pi$ is large enough for the snowball sample to cover most of the population network.
For a $1$-wave snowball sample ($r = 1$), the naive estimator is on average one to two orders of magnitude larger than the true value $\pi = 0.001$.
At $r = 2$, the naive estimator is still biased across all values of $\pi$, although the bias is reduced compared to $r = 1$.
Only at $r = 3$ and $\pi > 0.001$ does the naive estimator approach the true value of $\pi$, which corresponds to sampling more than $10{,}000$ out of $15{,}000$ vertices, as seen in Figure~\ref{fig:p_nodes}.

In contrast, the snowball-corrected estimator is approximately unbiased across all values of $\pi$ and $r$.
Notably, it remains accurate even with very sparse samples.
At $r = 1$ with fewer than $100$ sampled vertices on average, the average estimate is already close to the true $\pi$.
For all combinations of $\pi$ and $r$, the snowball-corrected estimator approaches the true value of $\pi$ more rapidly than the naive estimator as $r$ increases.

\section{Monte Carlo Confidence Intervals via Test Inversion}
\label{sec:mc_ci}

Having derived the corrected estimator $\hat\pi$ in Section~\ref{sec:MLE}, we now turn to quantifying its uncertainty.
A natural approach is to appeal to asymptotic normality and invert a Wald-type statistic built from the corresponding Fisher information.
Establishing such asymptotic normality, however, requires committing to a growth regime for the population network.
In particular, one needs to assume how the edge probability $\pi$ behaves as $N$ grows, i.e., an assumption about the sparsity of the network.
Because our data come from a single network of a fixed size, there is no principled way to justify one such assumption over another, and neither extreme is satisfactory.
If $\pi$ is held fixed, the network becomes dense so quickly that the second wave absorbs almost the entire remaining population and the snowball sample ceases to be a small subset of $V$, contrary to the setting motivating this paper.
If $\pi$ instead shrinks with $N$, the limiting behavior of $\hat{\pi}$ depends on the assumed rate, e.g. under $\pi = c/N$ for fixed $c > 0$, the size of the first wave, binomial by \eqref{eq:P-wave1}, converges to a Poisson distribution instead of concentrating around a growing mean, opposite of the regime in which a normal approximation would be justified.

To avoid this altogether, we instead construct confidence intervals for $\pi$ by Monte Carlo test inversion, inverting a level-$\alpha$ test of the hypothesis $\pi = \pi_0$ for every $\pi_0 \in (0,1)$ \citep{lehmann2022testing}.
The resulting confidence set is guaranteed to achieve at least the nominal coverage regardless of the values of $N$, $r$, and the underlying $\pi$.

To test the hypothesis $\pi = \pi_0$, we require a test statistic whose null distribution will be approximated.
A convenient and sensible choice is a function $h: \mathbb{R}^2 \to \mathbb{R}$ of the minimal sufficient statistic $(|E^{(r)}|, T(n))$, such as a likelihood-ratio statistic or the maximum likelihood estimator $\hat{\pi} = |E^{(r)}| / T(n)$ itself.
Let $h^{obs}$ denote the value of $h(|E^{(r)}|, T(n))$ at the observed $(|E^{(r)}|, T(n))$, and define the one-sided p-values
\begin{align*}
    p_L(\pi_0) &= P(h(|E^{(r)}|, T(n)) \geq h^{obs} \mid \pi_0), \\
    p_U(\pi_0) &= P(h(|E^{(r)}|, T(n)) \leq h^{obs} \mid \pi_0).
\end{align*}
Combine these into the two-sided p-value $q(\pi_0) = 2\min\{p_L(\pi_0), p_U(\pi_0)\}$, and reject $\pi = \pi_0$ whenever $q(\pi_0) \leq \alpha$ to obtain a level-$\alpha$ test.
Because $q(\pi_0) \leq \alpha$ if and only if $p_L(\pi_0) \leq \alpha/2$ or $p_U(\pi_0) \leq \alpha/2$, a union bound gives $P(q(\pi_0) \leq \alpha \mid \pi_0) \leq \alpha$, regardless of any dependence between $p_L(\pi_0)$ and $p_U(\pi_0)$.
By the duality between tests and confidence sets, inverting this test across every $\pi_0 \in (0,1)$ yields a conservative $100(1-\alpha)\%$ confidence set for $\pi$,
\begin{equation*}
    \{\pi_0 \in (0,1) : q(\pi_0) > \alpha\}.
\end{equation*}

As the null distribution of $h(|E^{(r)}|, T(n))$ often has no closed form, we approximate $p_L(\pi_0)$ and $p_U(\pi_0)$ by Monte Carlo simulation.
The $p$-values can be estimated by generating snowball samples under the candidate value $\pi_0$, using the algorithm of Section~\ref{subsec:er_simulation_verify}, and taking the proportion of simulated values of $h(|E^{(r)}|, T(n))$ exceeding, or falling below, $h^{obs}$.
Further computational gains are possible via importance sampling: a single batch of samples, simulated once under a reference value $\pi_Q$, can be reweighted to estimate $p$-values at any nearby candidate $\pi_0$ without resimulating.
Following \citet{harrison2012conservative}, the resulting $p$-value estimates $\hat p_L(\pi_0)$ and $\hat p_U(\pi_0)$ remain themselves valid, i.e., conservative $p$-values for the corresponding one-sided hypotheses.
The same union-bound argument therefore applies: given $\hat{q}(\pi_0) = 2\min\{\hat{p}_L(\pi_0), \hat{p}_U(\pi_0)\}$, rejecting $\pi = \pi_0$ whenever $\hat{q}(\pi_0) \leq \alpha$ defines a level-$\alpha$ test, and
\begin{equation}  \label{eq:conf_set}
    C = \{\pi_0 \in (0,1) : \hat{q}(\pi_0) > \alpha\}
\end{equation}
is a conservative $100(1-\alpha)\%$ confidence set for $\pi$.

If $\hat q(\pi_0)$ is suitably monotone or quasiconcave in $\pi_0$, the confidence set $C$ is an interval and can be located directly by the bisection search of \citet{glazer2026fast}.
Otherwise, we recover its convex hull $\operatorname{conv}(C) = [\inf C, \sup C]$ by searching inward from the edges of the parameter space, which retains the coverage guarantee established above.

To evaluate the procedure, we revisit the $9{,}000$ snowball samples generated for the point estimation study of Section~\ref{sec:MLE} ($N = 15{,}000$; $\pi \in \{0.001, 0.002, 0.003\}$; $r \in \{1,2,3\}$; $1{,}000$ replicates per combination).
We use the maximum likelihood estimator $\hat\pi = |E^{(r)}| / T(n)$ as the test statistic and construct a $95\%$ confidence interval $\operatorname{conv}(C)$ for each.
The exact simulation and parameter search implementation is given in Appendix~\ref{app:mc_ci_test}.
Table~\ref{tab:mc_ci_summary} reports the empirical coverage of $\operatorname{conv}(C)$, its Monte Carlo standard error $\sqrt{\widehat{\text{Coverage}}(1 - \widehat{\text{Coverage}}) / 1000}$, and its mean width for each $(\pi, r)$ combination.

\begin{table}[ht!]
    \centering
    \caption{Empirical coverage and width of the $95\%$ Monte Carlo confidence set $\operatorname{conv}(C)$, by true edge probability $\pi$ and number of waves $r$, over $1{,}000$ replicates per combination.}
    \label{tab:mc_ci_summary}
    \begin{tabular}{ccccc}
        \toprule
        $\pi$ & $r$ & Coverage & SE & Width \\
        \midrule
        0.001 & 1 & 0.960 & 0.006 & $1.08\times 10^{-3}$ \\
        0.001 & 2 & 0.950 & 0.007 & $2.57\times 10^{-4}$ \\
        0.001 & 3 & 0.944 & 0.007 & $4.89\times 10^{-5}$ \\
        0.002 & 1 & 0.952 & 0.007 & $1.45\times 10^{-3}$ \\
        0.002 & 2 & 0.951 & 0.007 & $2.01\times 10^{-4}$ \\
        0.002 & 3 & 0.962 & 0.006 & $2.10\times 10^{-5}$ \\
        0.003 & 1 & 0.953 & 0.007 & $1.73\times 10^{-3}$ \\
        0.003 & 2 & 0.943 & 0.007 & $1.44\times 10^{-4}$ \\
        0.003 & 3 & 0.948 & 0.007 & $2.05\times 10^{-5}$ \\
        \bottomrule
    \end{tabular}
\end{table}

Empirical coverage of $\operatorname{conv}(C)$ ranges from $0.943$ to $0.962$ across all nine combinations, with a mean of $0.951$.
Every cell lies within two Monte Carlo standard errors of the nominal level $0.95$, and none falls significantly below it, consistent with the coverage guarantee of at least $95\%$ for $C$ established in \eqref{eq:conf_set}.

The width of $\operatorname{conv}(C)$ shrinks by roughly an order of magnitude with each additional wave, reflecting the rapid growth of $T(n)$ with $r$.
For $r \geq 2$, it decreases with $\pi$, as larger $\pi$ produces larger waves and hence more informative samples, while the opposite holds for $r = 1$.

\section{Robustness and Real World Application}
\label{sec:robustness}

The results of Sections~\ref{sec:MLE} and~\ref{sec:mc_ci} establish that the snowball-corrected estimator is unbiased and its confidence intervals attain nominal coverage provided the population network is truly Erd\H{o}s--R\'enyi.
Real networks however are typically not, and a natural question is how the corrected estimator behaves once that model assumption fails.

First note that for an arbitrary random graph on $N$ vertices with $\pi_{i,j} = \mathbb{E}[Y_{i,j}]$, possibly heterogeneous across pairs and dependent in arbitrary fashion, the density of the full population network $\binom{N}{2}^{-1} |E|$ is unbiased for the average connection probability $\bar{\pi} = \binom{N}{2}^{-1} \sum_{\{i,j\} \in V^2} \pi_{i,j}$.

This unbiasedness result concerns an estimator computed on the full population network and is not guaranteed once only a partial, non-representative sample is available.
The naive density computed from the snowball sample, $\hat\pi_{\text{naive}}$, is biased for the average connection probability $\bar{\pi}$ under any model, ER included.
Similarly, the snowball-corrected estimator of Section~\ref{sec:MLE} is not expected to remain unbiased for non-ER population networks, as it divides the observed edge count by $T(n)$, a quantity derived from the wave-inclusion probabilities implied by the ER model.
Both estimators are thus affected by such misspecification, and both become unbiased as the sample approaches the full population, $U \to 0$.

Nevertheless, we still expect the snowball-corrected estimator to outperform the naive one, since it accounts for the exclusion of unsampled vertices from the sample, while the naive estimator does not.
We assess this improvement using synthetic networks, namely Barab\'asi--Albert and Watts--Strogatz models, alongside an application to a real Facebook ego network.

\subsection{Synthetic Networks}
\label{subsec:synthetic}

We consider two population models that depart from the ER assumption along different structural dimensions.
The Watts--Strogatz (WS) small-world model allows clustering to be tuned independently of degree, which remains homogeneous across vertices \citep{watts1998collective}.
The Barab\'asi--Albert (BA) model instead generates a heavy-tailed, heterogeneous degree sequence through preferential attachment, while retaining low clustering \citep{barabasi1999emergence}.

For both models, we match the simulation design of Section~\ref{sec:MLE}: $N = 15{,}000$ vertices, $r \in \{1,2,3\}$ waves, and $1{,}000$ replicates per combination.
Generative parameters are chosen to match the population densities $\pi \in \{0.001, 0.002, 0.003\}$ used for the ER simulations.
For BA, the average degree is set to $8$, $15$, and $23$, corresponding to densities of approximately $0.00107$, $0.00199$, and $0.00306$.
For WS, the number of neighbors per vertex is set to the same three values, with the rewiring probability fixed at $0.2$, yielding clustering coefficients between $0.18$ and $0.19$ \citep{barrat2000properties}, consistent with values observed in many real-world social and collaboration networks \citep{newman2018measures}.

Table~\ref{tab:ba_summary} and Table~\ref{tab:ws_summary} report, for each target density and number of waves, the average sampling fraction, the relative bias of the naive and snowball-corrected estimators, and the RMSE of the corrected estimator as a percentage of the naive estimator's RMSE.

\begin{table}[ht!]
    \centering
    \caption{Average sampling fraction, relative bias of the naive and snowball-corrected estimators, and RMSE of the snowball-corrected estimator as a percentage of the naive estimator's RMSE, under a Barab\'asi--Albert population, by target density $\pi$ and number of waves $r$, over $1{,}000$ replicates per combination. Bias and RMSE are computed relative to each replicate's realized density.}
    \label{tab:ba_summary}
    \begin{tabular}{ccccccc}
        \toprule
        $\pi$ & $r$ & Sampling frac. (\%) & Bias, naive (\%) & Bias, snow. (\%) & RMSE ratio (\%) \\
        \midrule
        0.001 & 1 & 0.11  & 15{,}071.85 & 0.32   & 0.56 \\
        0.001 & 2 & 3.80  & 833.02      & 172.99 & 21.85 \\
        0.001 & 3 & 57.35 & 56.72       & 42.37  & 73.20 \\
        0.002 & 1 & 0.21  & 4{,}630.10  & 7.34   & 1.83 \\
        0.002 & 2 & 11.34 & 303.74      & 180.36 & 58.12 \\
        0.002 & 3 & 96.14 & 3.27        & 2.69   & 84.25 \\
        0.003 & 1 & 0.31  & 2{,}200.70  & 14.81  & 3.70 \\
        0.003 & 2 & 22.33 & 167.23      & 136.83 & 80.44 \\
        0.003 & 3 & 99.96 & 0.03        & 0.03   & 78.15 \\
        \bottomrule
    \end{tabular}
\end{table}

\begin{table}[ht!]
    \centering
    \caption{Average sampling fraction, relative bias of the naive and snowball-corrected estimators, and RMSE of the snowball-corrected estimator as a percentage of the naive estimator's RMSE, under a Watts--Strogatz population, by target density $\pi$ and number of waves $r$, over $1{,}000$ replicates per combination. Bias and RMSE are computed relative to each replicate's realized density.}
    \label{tab:ws_summary}
    \begin{tabular}{ccccccc}
        \toprule
        $\pi$ & $r$ & Sampling frac. (\%) & Bias, naive (\%) & Bias, snow. (\%) & RMSE ratio (\%) \\
        \midrule
        0.001 & 1 & 0.11  & 26{,}481.36 & 134.03 & 0.54 \\
        0.001 & 2 & 1.21  & 2{,}349.40  & 43.91  & 1.97 \\
        0.001 & 3 & 13.24 & 183.91      & 26.70  & 14.48 \\
        0.002 & 1 & 0.21  & 12{,}117.45 & 264.66 & 2.23 \\
        0.002 & 2 & 3.89  & 526.29      & 69.06  & 13.29 \\
        0.002 & 3 & 57.86 & 13.58       & 3.43   & 27.20 \\
        0.003 & 1 & 0.31  & 7{,}332.82  & 398.52 & 5.49 \\
        0.003 & 2 & 8.62  & 190.39      & 62.17  & 32.73 \\
        0.003 & 3 & 94.87 & 0.56        & $-0.39$ & 65.15 \\
        \bottomrule
    \end{tabular}
\end{table}

Across both models and all nine combinations of $\pi$ and $r$, the snowball-corrected estimator attains a strictly lower RMSE than the naive estimator, with the ratio never exceeding $85\%$.
The naive estimator's relative bias is extreme throughout, exceeding $15{,}000\%$ under BA and $26{,}000\%$ under WS at $r=1$.
The corrected estimator's bias, while remaining far smaller, is still considerable reaching at most $398\%$.
In both models, bias falls toward zero as the sampling fraction approaches $100\%$, since the two estimators converge to the unbiased population density estimator with $U \to 0$.

While the naive estimator's bias decreases monotonically with the number of waves across the two models, the corrected estimator's bias behaves differently.
Under BA, it falls, rises, then falls again: at $r=1$ the seed vertex is drawn uniformly and is unlikely to be a hub, so the sample locally resembles an ER graph and bias is small; at $r=2$ preferential attachment makes the newly recruited wave disproportionately likely to include high-degree vertices, causing a jump in bias; and at $r=3$ the sampling fraction is high enough that both estimators converge regardless.
Under WS, the corrected estimator's bias decreases monotonically with the number of waves, as its clustering coefficient remains elevated relative to a matched-density ER network throughout.

\subsection{Facebook Ego Network}
\label{subsec:facebook}

In practice, researchers rarely have direct access to a large online social network in its entirety.
Platform application programming interfaces and privacy restrictions commonly limit access to ego networks centered on a small number of seeds.
An ego network is constructed by choosing a predetermined ego vertex, including all of its direct neighbors (the alters), and recording all edges within alters and between the alters and the ego.
It is thus equivalent to a $1$-wave snowball sample.

We illustrate how much can nonetheless be recovered about the underlying population network from such limited access, estimating the edge density of the entire Facebook network from a single ego network, treating the global network as an ER population.
The ego network is taken from the Facebook friendship dataset collected by \citet{maier2017cover}, obtained from one author's Facebook friend list along with the connections among those friends, as recorded in April 2014\footnote{Data available at \url{https://github.com/benmaier/BFMaierFBnetwork}.}.
The sample comprises $n_1 = 362$ alters and $|E^{(r)}| = 2350$ edges in total.
We take the population size $N$ to be the number of active Facebook users at the time of data collection, reported as $1.28$ billion monthly active users in Meta's first-quarter 2014 operational highlights \citep{facebook2014q1}.

\begin{figure}[!ht]
    \centering
    \includegraphics[scale = 0.65]{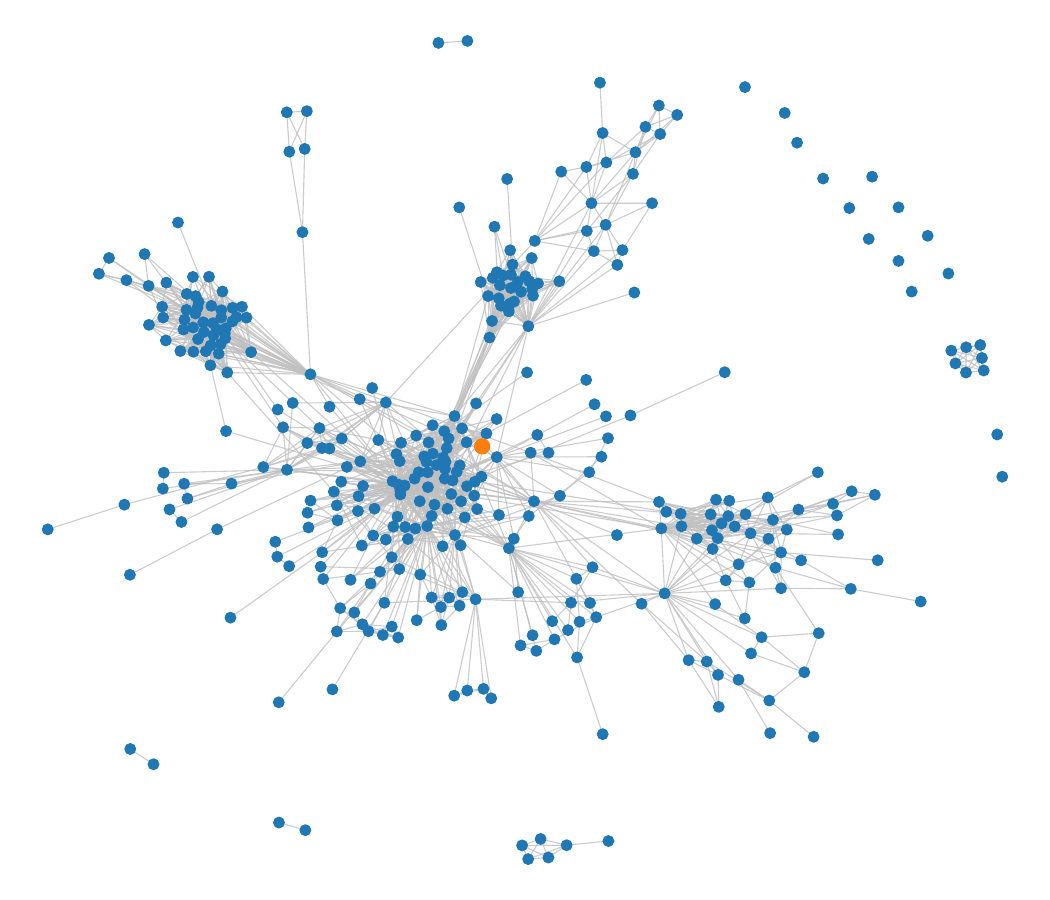}
    \caption{The Facebook ego network of \citet{maier2017cover}. The ego vertex is shown in orange, and all edges connecting to the ego are omitted.}
    \label{fig:facebook_ego}
\end{figure}

Figure~\ref{fig:facebook_ego} shows the resulting ego network, with the ego shown separately in the center for reference, its edges omitted for readability.

Applying the naive estimator \eqref{eq:er_naive_mle} and the snowball-corrected estimator \eqref{eq:er_snow_mle} to the Facebook ego network yields a snowball-corrected estimate of $\hat{\pi} = 1.835 \cdot 10^{-6}$, with a $95\%$ Monte Carlo confidence interval of $[1.768 \cdot 10^{-6}, 1.910 \cdot 10^{-6}]$, whereas the naive estimate is $\hat{\pi}_{\text{naive}} = 0.0358$.
Under the naive estimator, the implied population-average degree is $\hat{\pi}_{\text{naive}}N \approx 45.8 \cdot 10^6$, exceeding the $5{,}000$-friend limit Facebook enforces per personal profile \citep{facebook2014friendlimit} by more than a factor of $9{,}000$; no single account could have anywhere near this many connections, so the naive estimate cannot plausibly describe the population-wide density.
The snowball-corrected estimator instead implies an average degree of $\hat{\pi} N \approx 2{,}349$.
This is markedly more plausible, though still roughly an order of magnitude above contemporaneous survey figures.
A Pew Research survey conducted in February 2014 found a mean of $338$ and a median of $200$ friends among adult Facebook users in the United States \citep{pew2014facebook}.

This gap is consistent with the pattern established in Section~\ref{subsec:synthetic}.
Real social networks exhibit both degree heterogeneity and clustering, and the corrected estimator, though it substantially outperforms the naive one, is not expected to be unbiased once the population departs from the ER assumption.
The same caveat extends to the coverage of the Monte Carlo confidence interval, which was established under a correctly specified model.

\section{Discussion}
\label{sec:discussion}
This paper derives the exact likelihood of a multi-wave snowball sample under the Erdős–Rényi model, resolving the base case of Research Problem 3.1 posed by \citet{crane2018network}, and uses it to construct a snowball-corrected estimator of the population's edge density along with confidence intervals obtained by inverting a test based on the exact sampling distribution.
Simulations show that the naive estimator, which ignores the sampling design, can overstate density by one to two orders of magnitude for sparse networks observed over a single wave, while the corrected estimator remains approximately unbiased even when the sample covers a small fraction of the population, and the confidence intervals attain nominal coverage while remaining substantially narrower than an uncorrected bracket.
Applied to a real Facebook ego network, the corrected estimator implies a population-average degree several orders of magnitude more plausible than the naive estimator's.

This last result, however, comes with a caveat central to the paper's framing: the equivalence between the model parameter $\pi$ and the population density holds only under the Erdős–Rényi assumption, and Section~\ref{sec:robustness} shows that this equivalence does not survive departures from it.
When the population instead exhibits degree heterogeneity, clustering, or community structure, the snowball-corrected estimator no longer recovers the true density, and in the Facebook application the population-average degree implied by the fitted model differs from independently reported survey figures by roughly a factor of seven.
The method should therefore be read as an exact density estimator under a fully specified model, and as a benchmark for assessing sensitivity to that specification, rather than as a model-free tool.

A natural extension is to models that preserve dyad independence conditional on latent vertex-level quantities, such as latent space \citep{hoff2002latent}, stochastic block \citep{holland1983stochastic}, or graphon models \citep{lovasz2006limits}: because these retain the structure that makes the Erdős–Rényi case tractable, we expect the same likelihood correction to extend to them with comparatively modest modification. This is left to future work.

Two further limitations are worth noting briefly. First, our derivation assumes the ego $v_0$ is selected independently of the network, as in the Facebook application; endogenous ego selection would make the selection event informative about $\pi$ and require a selection probability that depends on the unsampled graph, so a closed-form correction is not immediate. Second, an argument can be made about whether or not edges between nodes in the last observed snowball wave should be treated as observed or not. In practice, revealing wave $r-1$'s neighbors exposes wave $r$ and the edges connecting it to wave $r-1$, but not the edges within wave $r$ itself. In Appendix~\ref{app:last_wave}, we show that this non-observability of edges within the last wave does not change our results and resolves cleanly with minor adjustments.

Taken together, these results show that the sampling mechanism inherent to snowball designs can be incorporated directly into the likelihood at little additional cost whenever the underlying network model factorizes over dyads, yielding a corrected density estimator and valid confidence sets grounded in the exact sampling distribution rather than an asymptotic approximation — while also showing plainly why that correction is only as trustworthy as the population model it assumes.

\bibliography{bibliography}

\appendix
%\section{Endogenous Ego Selection}
%\label{subsec:endogenous_ego}

\section{Computational Details for the Monte Carlo Confidence Set \label{app:mc_ci_test}}
This appendix collects the computational details deferred from Section~\ref{sec:mc_ci}: the Monte Carlo estimation of $p_L(\pi_0)$ and $p_U(\pi_0)$ via importance sampling, the procedure for locating the confidence set $C$ and its convex hull, and the simulation settings used in the empirical study.

For each simulated sample, write $(|E_j^{(r)}|, T(n)_j)$, $j = 1, \dots, M$, for its sufficient statistic under the proposal $\pi_Q$, discarding any draw with $T(n)_j = 0$, which carries no information about $\pi$.
Following \eqref{eq:exp_family_form}, the batch can be reweighted to represent a sample drawn under any other value $\pi_0 \in (0,1)$ via the importance weight
\begin{equation*}
    w_j(\pi_0) = \frac{\pi_0^{|E_j^{(r)}|}(1-\pi_0)^{T(n)_j-|E_j^{(r)}|}}{\pi_Q^{|E_j^{(r)}|}(1-\pi_Q)^{T(n)_j-|E_j^{(r)}|}},
\end{equation*}
which is well defined for every $\pi_0 \in (0,1)$ since the support $\mathcal{S}(n)$ of the sampling distribution does not depend on $\pi$ (Section~\ref{subsec:cond_prob}).
Let $w^{obs}(\pi_0)$ denote the importance weight evaluated at the observed $(|E^{(r)}|, T(n))$.
The importance-sampling p-value estimates of \citet{harrison2012conservative} are
\begin{align*}
    \hat{p}_L(\pi_0) &= \frac{w^{obs}(\pi_0) + \sum_{j=1}^M w_j(\pi_0)\,\mathds{1}(h(|E_j^{(r)}|, T(n)_j) \geq h^{obs})}{w^{obs}(\pi_0) + \sum_{j=1}^M w_j(\pi_0)}, \\
    \hat{p}_U(\pi_0) &= \frac{w^{obs}(\pi_0) + \sum_{j=1}^M w_j(\pi_0)\,\mathds{1}(h(|E_j^{(r)}|, T(n)_j) \leq h^{obs})}{w^{obs}(\pi_0) + \sum_{j=1}^M w_j(\pi_0)}.
\end{align*}
Direct Monte Carlo simulation at $\pi_0$ is the special case $\pi_Q = \pi_0$, where every weight equals one and $\hat p_L(\pi_0), \hat p_U(\pi_0)$ reduce to the usual add-one Monte Carlo p-value estimators.
Following \citet{harrison2012conservative}, $\hat{p}_L(\pi_0)$ and $\hat{p}_U(\pi_0)$ satisfy $P(\hat{p}_L(\pi_0) \leq p \mid \pi_0) \leq p$ and $P(\hat{p}_U(\pi_0) \leq p \mid \pi_0) \leq p$ for all $p \in [0,1]$, for any $M \geq 1$ and any proposal $\pi_Q$.
Combining these into the two-sided estimate $\hat q(\pi_0) = 2\min\{\hat p_L(\pi_0), \hat p_U(\pi_0)\}$ and inverting the resulting test, as in Section~\ref{sec:mc_ci}, yields the confidence set $C$ in \eqref{eq:conf_set}.

\citet{glazer2026fast} propose a modified bisection search to construct such confidence sets efficiently when the p-value function is monotone or quasiconcave in the parameter.
When this assumption is violated, however, the level set \eqref{eq:conf_set} need not be an interval, and bisection is not guaranteed to recover it correctly.
We therefore avoid assuming quasiconcavity and instead target the convex hull of the confidence set,
\begin{equation*}
    \operatorname{conv}(C) = [\inf C, \sup C].
\end{equation*}
Since $C \subseteq \operatorname{conv}(C)$, the coverage guarantee established for $C$ in Section~\ref{sec:mc_ci} carries over to its convex hull: $P(\pi \in C) \geq 1 - \alpha$ implies $P(\pi \in \operatorname{conv}(C)) \geq 1 - \alpha$ for every $\pi \in (0,1)$.
When $\hat{q}(\pi_0)$ is in fact quasiconcave, $C$ is already an interval and $\operatorname{conv}(C) = C$.
Otherwise, $\operatorname{conv}(C)$ remains a valid, if potentially conservative, confidence set.

Because $\operatorname{conv}(C)$ is an interval, it suffices to locate its two endpoints, $\inf C$ and $\sup C$.
Rather than searching the full $(0,1)$ range, we first obtain a cheap starting bracket from the relative likelihood region for $\pi$, i.e., the set of parameter values whose log-likelihood \eqref{eq:log_like} lies within a threshold of its maximum,
\begin{equation*}
    \{\pi_0 : \ell(\pi_0) - \ell(\hat\pi^{obs}) > \log \kappa\},
\end{equation*}
for a cutoff $\kappa \in (0,1)$ \citep{pawitan2001elements}.
Since $\ell$ is strictly concave in $\pi_0$ for fixed $(|E^{(r)}|, T(n))$ with unique maximum at $\hat\pi^{obs}$, this region is an interval $[\pi_0^{lo}, \pi_0^{hi}]$, whose endpoints can each be located by bisection.
Because it treats $(|E^{(r)}|, T(n))$ as fixed, ignoring the randomness of $T(n)$ under $\pi_0$, this interval is not itself a valid confidence set for $\pi$ in the sense of \eqref{eq:conf_set}.
We use it only as a computational starting bracket for the marching search below.
For $\kappa$ small, $[\pi_0^{lo}, \pi_0^{hi}]$ is comfortably wider than $\operatorname{conv}(C)$ while remaining far narrower than $(0,1)$, giving a much closer starting point than searching the full parameter space.

We locate each endpoint of $\operatorname{conv}(C)$ by marching inward from the corresponding bracket edge toward $\hat\pi^{obs}$ in steps of $\epsilon$ until $\hat q(\pi_0)$ first exceeds $\alpha$, then refining that crossing by bisection to tolerance $\tau$.
Each evaluation of $\hat q(\pi_0)$ along the march reweights the current batch via $w_j(\pi_0)$, and because $T(n)_j$ can range from a few thousand to several million across the candidate values visited, these weights degenerate sharply once $\pi_0$ drifts far from the proposal $\pi_Q$, concentrating on a handful of the $M$ draws and making $\hat q(\pi_0)$ unreliable despite $M$ being nominally large.
We monitor this via the effective sample size of the importance weights,
\begin{equation*}
    \text{ESS}(\pi_0) = \frac{\big(\sum_{j=1}^M w_j(\pi_0)\big)^2}{\sum_{j=1}^M w_j(\pi_0)^2} \in [1, M],
\end{equation*}
redrawing a fresh batch under $\pi_Q = \pi_0$ whenever $\text{ESS}(\pi_0)$ falls below a fixed floor, and reusing it for subsequent evaluations until it too degenerates.
Because consecutive points along the march lie close together, most evaluations reuse the current batch at negligible cost.
Since $\hat q(\pi_0)$ is continuous in $\pi_0$ for any fixed batch, the resulting estimates of $\inf C$ and $\sup C$ converge to their true values as $\epsilon$ and $\tau$ tend to zero.

For the study reported in Section~\ref{sec:mc_ci}, each confidence set is computed from an initial batch of $M = 1{,}500$ importance samples, redrawn whenever $\text{ESS}(\pi_0)$ falls below $1{,}000$.
The march step is set to $\epsilon = 10^{-6}$, the bisection tolerance to $\tau = 10^{-8}$, and the relative likelihood cutoff to $\kappa = 0.001$.

For comparison, Table~\ref{tab:mc_ci_rl} reports the same coverage and width of $\operatorname{conv}(C)$ from Table~\ref{tab:mc_ci_summary} alongside the containment rate and mean width of the relative likelihood bracket $[\pi_0^{lo}, \pi_0^{hi}]$ used to initialize each march.
\begin{table}[ht!]
    \centering
    \caption{Comparison between the confidence set $\operatorname{conv}(C)$ and the $99.9\%$ relative likelihood (RL) starting bracket used to initialize its search, reporting the coverage of each and the reduction in width of $\operatorname{conv}(C)$ relative to the RL bracket, by true edge probability $\pi$ and number of waves $r$, over $1{,}000$ replicates per combination.}
    \label{tab:mc_ci_rl}
    \begin{tabular}{ccccc}
        \toprule
        $\pi$ & $r$ & Coverage & RL bracket cont. & Width reduction \\
        \midrule
        0.001 & 1 & 0.960 & 1.000 & 44.4\% \\
        0.001 & 2 & 0.950 & 1.000 & 43.9\% \\
        0.001 & 3 & 0.944 & 1.000 & 41.5\% \\
        0.002 & 1 & 0.952 & 0.999 & 45.9\% \\
        0.002 & 2 & 0.951 & 0.999 & 44.7\% \\
        0.002 & 3 & 0.962 & 0.999 & 43.9\% \\
        0.003 & 1 & 0.953 & 0.999 & 46.3\% \\
        0.003 & 2 & 0.943 & 1.000 & 44.5\% \\
        0.003 & 3 & 0.948 & 1.000 & 46.6\% \\
        \bottomrule
    \end{tabular}
\end{table}
The confidence interval $\operatorname{conv}(C)$ is on average $44.6\%$ narrower in comparison, ranging between $41.5$ and $46.6\%$ across the nine combinations, which demonstrates that the marching-and-bisection refinement yields a substantially tighter set than the starting bracket alone while retaining the conservative coverage guarantee.

\section{Observability of Edges Within the Last Wave \label{app:last_wave}}

In practice, revealing the neighbors of the vertices in wave $r$ also reveals the vertices of wave $r+1$ and the edges connecting them to wave $r$, but not the edges within wave $r+1$ itself, as exposing those would in turn reveal wave $r + 2$, and so on.
This raises the question whether the edges within the last observed wave, $r$, should be treated as observed at all.
We show that omitting them poses no difficulty for our framework.

Let $W_r = \{\{i,j\} \in W : i, j \in V^{(r)}\}$ denote the vertex pairs within the last wave, and $W_{-r} = W \setminus W_r$ the remaining within-wave pairs.
The marginal probability of the wave sets \eqref{eq:wave_marginal} depends only on wave inclusion and is unaffected by whether $\ry_{W_r}$ is observed.
Because the edge indicators in $W$ are independent and identically distributed Bernoulli variables unconstrained by the snowball design, $P(\ry_W) = P(\ry_{W_r})P(\ry_{W_{-r}})$ and so \eqref{eq:conditional_mat} can be written as
\begin{equation*}
    P(\ry \mid V^{(0)}, \dots, V^{(r)}) = P(\ry_A \mid V^{(0)}, \dots, V^{(r)}) \, P(\ry_{W_{-r}}) \, P(\ry_{W_r}).
\end{equation*}
Summing over the $2^{|W_r|}$ possible configurations of $\ry_{W_r}$ yields
\begin{align*}
    \sum_{\ry_{W_r} \in \{0,1\}^{|W_r|}} P(\ry \mid V^{(0)}, \dots, V^{(r)}) &= P(\ry_A \mid V^{(0)}, \dots, V^{(r)}) \, P(\ry_{W_{-r}}) \sum_{\ry_{W_r} \in \{0,1\}^{|W_r|}} P(\ry_{W_r}) \\
    &= P(\ry_A \mid V^{(0)}, \dots, V^{(r)}) \, P(\ry_{W_{-r}}),
\end{align*}
since $\sum_{\ry_{W_r} \in \{0,1\}^{|W_r|}} P(\ry_{W_r}) = 1$ as $\ry_{W_r}$ is a vector of independent and identically distributed Bernoulli variables.
Therefore,
\begin{equation*}
    P(\ry_A, \ry_{W_{-r}} \mid V^{(0)}, \dots, V^{(r)}) = \frac{\prod_{\{i,j\} \in A \cup W_{-r}} \pi^{\ry_{i,j}}(1 - \pi)^{1 - \ry_{i,j}}}{\prod_{k = 1}^r [1 - (1 - \pi)^{n_{k-1}}]^{n_k}}.
\end{equation*}
Following the same steps as Section~\ref{sec:erdos}, the joint probability of the wave sets and the observable adjacency matrix is
\begin{equation*}
    P(\ry, V^{(1)}, \dots, V^{(r)} \mid V^{(0)}) = \pi^{|E^{(r)}_{-r}|} (1-\pi)^{T(n) - |W_r| - |E^{(r)}_{-r}|},
\end{equation*}
where $|E^{(r)}_{-r}| = \sum_{\{i,j\} \in A \cup W_{-r}} \ry_{i,j}$.
The minimal sufficient statistic is therefore $(|E^{(r)}_{-r}|,\, T(n) - |W_r|)$, and the corresponding maximum likelihood estimator is
\begin{equation*}
    \hat{\pi}_{-r} = \frac{|E^{(r)}_{-r}|}{T(n) - |W_r|}.
\end{equation*}
Thus, not observing edges within the last wave is formally equivalent to treating the $|W_r|$ pairs as unsampled. They are simply dropped from both the numerator and denominator of \eqref{eq:er_snow_mle}, exactly as if they had never entered the sample.

\bibliographystyle{chicago}

\end{document}